\documentclass[12pt,nohyper,notoc]{JHEP}

\usepackage{amsmath,amssymb,cite,graphicx} 
	
\newcommand{\beq}{\begin{eqnarray}}
\newcommand{\eeq}{\end{eqnarray}}
\newcommand{\drawsquare}[2]{\hbox{%
\rule{#2pt}{#1pt}\hskip-#2pt
\rule{#1pt}{#2pt}\hskip-#1pt
\rule[#1pt]{#1pt}{#2pt}}\rule[#1pt]{#2pt}{#2pt}\hskip-#2pt
\rule{#2pt}{#1pt}}

\newcommand{\Yfund}{\raisebox{-.5pt}{\drawsquare{6.5}{0.4}}}

\font\zfont = cmss10 
\newcommand\ZZ{\hbox{\zfont Z\kern-.4emZ}}
\def\inbar{\vrule height1.5ex width.4pt depth0pt}
\def\IC{\relax\hbox{\kern.25em$\inbar\kern-.3em{\rm C}$}}

\newcommand{\EQ}[1]{\begin{equation} #1 \end{equation}}

\title{Holographic RG and Cosmology in Theories with 
Quasi-Localized Gravity}

\author{Csaba Cs\'aki$^{a,}$\footnote{J. Robert Oppenheimer Fellow.}, 
Joshua Erlich$^a$, Timothy J. Hollowood$^{a,b}$ and John Terning$^c$\\
$^a$Theory Division T-8, Los Alamos National Laboratory, Los Alamos,
NM 87545, USA\\
$^b$Department of Physics, University of Wales Swansea,
Swansea, SA2 8PP, UK\\
$^c$Department of Physics, Harvard University,
Cambridge, MA 02138, USA\\

Email: {\tt csaki@lanl.gov, erlich@lanl.gov, pyth@skye.lanl.gov,
terning@schwinger.harvard.edu}}

\abstract{
We study the long distance behaviour of brane theories with 
quasi-localized gravity.  The 5D effective theory at large scales follows from
a holographic renormalization group flow.  As intuitively
expected, the graviton  is effectively four dimensional at intermediate 
scales and becomes five dimensional
at large scales.  However in the holographic effective theory the essentially
4D radion dominates at long distances and gives
rise to scalar anti-gravity. The holographic description shows
that at large distances the GRS model is equivalent to the 
model recently proposed by Dvali, Gabadadze and Porrati (DGP), where a 
tensionless brane is embedded into 5D Minkowski space, with an additional
induced 4D Einstein-Hilbert term on the brane. In the holographic description
the radion of the GRS
model is automatically localized on the tensionless brane, and provides the
ghost-like field necessary to cancel the extra graviton polarization of the
DGP model.  Thus, there is a holographic duality between these theories.
This analysis provides  physical insight into how the 
GRS model works at intermediate scales; 
in particular it sheds light on the size of the
width of the graviton resonance, 
and also demonstrates how the holographic RG can be used as a practical tool
for calculations.
}

\preprint{{\tt hep-th/0003076} \\ HUTP-00/A008}

\begin{document}

\section{Introduction}

Randall and Sundrum (RS) have recently shown that it is possible to localize
gravity to a brane in five dimensional anti-de Sitter space
\cite{RS,RS2}. In this model
the theory on the brane reproduces four dimensional Einstein gravity
at large distances even though the size of the extra dimension is infinitely
large. This idea has sparked a flurry of activity in this field \cite{Nima,CS,
junction,cvetic,Gremm,US,Gremm2,Linde,skenderis,Kostas,CHR,GT,GKR,
Japanese,CGS2,other} 
(see \cite{noncompact,oldCvetic} for related earlier work). 
Gregory, Rubakov and Sibiryakov (GRS) \cite{GRS}
proposed a modified version of the RS model in which gravity appears 
five dimensional both at short and at long distances (see \cite{Kogan} for
a related idea), while at intermediate scales four dimensional gravity
is reproduced. The GRS model has a positive tension brane, as in the RS model, and 
negative
tension branes at a large proper distance away from
the positive tension brane on either side of it. These negative tension 
branes have half the tension of the central brane.
The brane tensions and the negative
bulk cosmological constant are tuned such that the background is
static, and the geometry between the branes is a slice of AdS$_5$, while beyond
the negative tension branes the spacetime is ordinary 5D
Minkowski space. The fact that 4D gravity is reproduced at intermediate
scales  has been explained in \cite{GRS,CEH,Dvali}, and the 
reason for this is that in this model the 4D graviton is replaced by 
a resonance with a finite lifetime, which can decay into the bulk;
thus gravity is only quasi-localized. It was also
suggested in \cite{Dvali,Witten} that there may be a connection in these
theories to bulk supersymmetry and vanishing of the cosmological
constant.
It has been shown in \cite{CEH2,GRS2}, 
that at intermediate distance scales the theory indeed reproduces the results
of ordinary general relativity due to the bending of the brane in the presence
of inhomogenous matter on the brane. The reason for this is that the effect
of the bending of the brane exactly cancels the effects of the extra 
polarization in the massive graviton propagator (up to corrections that
can be made arbitrary small by adjusting the width of the resonance).  

We should stress that one of the essential features of these models is that
they do not have a 4D low-energy effective field theory description (see also
\cite{Witten}). Instead of an effective 4D theory, these models have a 
an effective 5D theory at large distances
which can be derived through a holographic 
renormalization group (RG) flow. 
This holographically renormalized  theory
will play the role of the low-energy effective theory. 
 In order to simply perform a calculation at a 
given energy scale on the
brane one  performs the RG running to that 
scale in the theory on the brane. From the 
conjectured AdS/CFT correspondence the
RG flow corresponds to
moving the brane a finite distance into the
bulk. This procedure is referred to as the 
holographic RG  
and will be the key to understanding the theory at
large distances. At intermediate energies, the RG flow corresponds to moving
the brane inside the AdS slice. Since an exactly AdS bulk corresponds to a
conformal field theory, the brane tension 
of the effective theory remain unchanged. However, at 
distances 
large
enough that effectively the branes have crossed, the effective low-energy 
theory will be that of a 
tensionless brane in 5D Minkowski space. This procedure can also be 
implemented for models which are smooth versions of the GRS model 
(see \cite{CEH}), with the difference that there will be a continuous 
running in the brane tension of the theory. However, as long as the asymptotic 
metrics are equivalent to those of GRS the asymptotic form of the low energy 
effective  theory will be the same as for GRS.
In fact, a detailed analysis below will show that one finds
that at intermediate distances (more precisely at a scale $k^{-1} e^{ky_0}$,
where $y_0$ is the location of the negative tension brane in the GRS model)
the model becomes equivalent to the model recently proposed by Dvali,
Gabadaze and Porrati \cite{DGP3}, where a tensionless brane is embedded
into 5D Minkowski space, but there is an additional induced four
dimensional curvature term on the brane present. The induced operator
is a consequence of the holographic renormalization. In addition, the
radion mode of the GRS model (which corresponds to fluctuations of the 
distance between the two branes) will also be localized on the tensionless
brane, as predicted in \cite{PRZ}. This radion field will have a wrong-signed
kinetic term, which is needed to cancel the effects of the extra graviton
polarization in the DGP model. However, at very long distances, where the
graviton mode becomes 5 dimensional, the radion will start to dominate and 
give rise to a peculiar 4D scalar antigravity, as discussed in 
\cite{GRS2,PRZ}. Due to the negative kinetic term of the radion
these theories are probably not internally consistent
at large scales; but from a purely
phenomenological point of view it is still interesting
to use these results to see how the cosmology 
of these models deviates from the ordinary FRW expansion of the Universe
at large scales.

The paper is organized as follows: 
in Section 2 we explain the basic idea behind the holographic 
renormalization group, and calculate the effective brane tension 
and induced curvature term on the brane. In Section 3 we review the
calculation of  the induced
radion kinetic term on the effective brane. In Section 4 we use the
effective holographic theory obtained in Sections 2 and 3 to calculate the
graviton propagator at large distances. We speculate on the cosmology
of these models in Section 5, and conclude in Section 6.

\section{Holographic Renormalization in Quasi-Localized Gravity Scenarios}

Consider a 5D metric of the form
\EQ{
ds^2=dy^2+e^{-A(y)}\eta_{\mu\nu}dx^\mu\,dx^\nu\ ,
\label{nmetric}
}
where the warp factor $A(y)$ approaches the
AdS form for $|y|\ll y_0$:
\EQ{
A(y)\to 2k|y| \ ,
}
while for $|y|\gg y_0$ the metric becomes flat:
\EQ{
A(y)\to {\rm constant}\ .
}
In the GRS model, \cite{GRS} this is achieved by simply
patching AdS$_5$ to flat space at some point $y=y_0$:
\EQ{
e^{-A(y)}=\begin{cases} e^{-2k|y|} & |y|\leq y_0\\
e^{-2ky_0} & |y|\geq y_0\ .
\end{cases}
}
We can also construct examples which smoothly interpolate between
AdS$_5$ and flat space \cite{CEH}:
\EQ{
e^{-A(y)}=e^{-2k|y|}+e^{-2ky_0}\ .
\label{smm}
}
In this case the cross-over from AdS to flat space occurs over a small
region $\delta y\sim k^{-1}$. In both these scenarios we need
$e^{ky_0}\gg1$ so that the cross-over to flat space occurs at a large
proper distance from the origin in the transverse space.
Since the metric approaches the RS metric for $|y|\ll y_0$, there is a
brane located at $y=0$ with a
positive tension $V=6k/\kappa^2$ (the ``Planck brane'') 
on which the matter fields will live. In the GRS model, there
are additional branes at $y=\pm y_0$ with negative tensions
$-\tfrac12 V$. In the smoothed version \eqref{smm}, the region of
negative tension is smeared over a scale $\delta y\sim k^{-1}$. As pointed out
in 
\cite{CEH,Witten} such smooth backgrounds violate positivity, which will
give rise to the instability discussed in \cite{PRZ}.

In \cite{CEH2}, the propagator for gravity on
the positive tension Planck brane at $y=0$ was found to have the form
\EQ{
G(x,x')_{\mu\nu,\rho\sigma}=
\Delta_5(x,0;x',0)\big(\tfrac12\eta_{\mu\rho}\eta_{\nu\sigma}
+\tfrac12\eta_{\mu\sigma}\eta_{\nu\rho}-\tfrac13\eta_{\mu\nu}
\eta_{\rho\sigma}\big)-\tfrac k6\Delta_4(x,x')\eta_{\mu\nu}
\eta_{\rho\sigma}\ .
\label{gprop}
}
Here, the first term is what would na\"\i vely be expected, since the
tensor structure is five dimensional, and $\Delta_5(x,y;x',y')$ is the
scalar Green's function for the background \eqref{nmetric}:
\EQ{
\big(e^A\square^{(4)}+\partial_y^2-2A^{\prime}\partial_y\big)\Delta_5(x,y;x',y')
=e^{2A} \delta^{(4)}(x-x')\delta(y-y')\ ,
}
The unexpected piece is the last term in (\ref{gprop}),
which occurs because matter
sources on the brane actually bend the brane, and this effect modifies the
braneworld propagator. Notice that the brane-bending term involves the
four-dimensional massless scalar propagator defined via
\EQ{
\square^{(4)}\Delta_4(x,x')=\delta^{(4)}(x-x')\ .
}
If the negative tension branes are sufficiently far away
($e^{ky_0}\gg1$), then there is a large intermediate region of
four-dimensional distance scales $k^{-1}\ll r\ll k^{-1}e^{3ky_0}$,
where $\Delta_5(x,0;x',0)$ is dominated by a zero
energy resonance and is approximately $k\Delta_4(x,x')$. 
In this
region the two terms in \eqref{gprop} combine into the usual 4D
graviton propagator:
\EQ{
G(x,x')_{\mu\nu,\rho\sigma}\to kG_4(x,x')_{\mu\nu,\rho\sigma}=
k\Delta_4(x,x')\big(\tfrac12\eta_{\mu\rho}\eta_{\nu\sigma}
+\tfrac12\eta_{\mu\sigma}\eta_{\nu\rho}-\tfrac12\eta_{\mu\nu}
\eta_{\rho\sigma}\big)\ .
}
Hence, at these intermediate distance scales conventional gravity is recovered as a
consequence of the interplay between the resonance and the bending of
the brane. This makes intuitive
sense: suppose we send out signals ({\it e.g.\/} a gravitational wave pulse)
from a point on the brane, along
the brane and transverse to it. If the signal along the brane
only travels a distance  $r \ll k^{-1}e^{ky_0}$, in the same proper time
the transverse signal only
explores the AdS portion of the transverse space and we expect that the physics
on the brane is unchanged from the RS scenario where the AdS space extends
out to infinite $y$ \cite{RS}. 

At larger distance scales on the brane the corresponding
transverse signal reaches the flat portion of the transverse space and
so the physics at scales 
$r \gg  k^{-1}e^{ky_0}$ on the brane will be modified from
pure 4D gravity. In fact there is a cross-over region
at distance scales $r\simeq k^{-1}e^{3ky_0}$, where
$\Delta_5(x,0;,x',0)$ becomes the massless scalar
propagator in 5D Minkowski space. So at large distances the first term
in \eqref{gprop} gives rise to a gravitational potential that falls off as
$1/r^2$. However, the brane-bending term is always four dimensional and gives 
rise to a gravitational potential that falls
off as
$1/r$. Hence at very large distances the brane-bending
term dominates and gives rise to scalar anti-gravity as noted in \cite{GRS2}. 

The same result has been also obtained by Pilo, Rattazzi and Zaffaroni (PRZ)
\cite{PRZ}, who have calculated the graviton propagator on the brane by
identifying all relevant physical modes of the theory. These were 
shown to be  the graviton resonance, and the massless radion field describing 
the
fluctuations of the distance between the two branes, which turns out to
have a negative kinetic term. The interpretation of the brane bending
calculation summarized above is then that at intermediate energies, where the
graviton is effectively localized, the negative kinetic term radion
exactly cancels the contribution of the extra polarization in the massive
graviton propagator, thus reproducing 4D Einstein gravity. For large distances
however, the radion will dominate, and thus lead to the scalar antigravity
as predicted in \cite{GRS2}.

Below we will show how to use the technology known as the holographic
renormalization group to obtain these results. In the meantime we
clarify the meaning of holographic renormalization. This procedure
corresponds to a renormalization group {\it coarse graining\/} which 
simply describes the
effective physics on the brane. As usual, this involves
integrating out degrees-of-freedom up to a certain physical length scale on
the brane, which we denote as $\tilde r$, but which now also includes an 
averaging over the transverse space out to proper distances $\tilde r$. 
A proper
distance $r$ on the brane corresponds to going out to $y$ in the
transverse direction, where
\EQ{
r=\int^{y}_0e^{A(y')/2}\,dy'\ .
\label{distance}}
This can be understood in two different ways. First, one can  calculate
how far into the bulk light can travel, while traveling a distance 
$r$ on the brane. For light $ds^2=0$, which yields $dr=e^{A(y)/2} dy$, thus
yielding (\ref{distance}). Another way of obtaining the correspondence
between distances on the brane and the bulk is to check how far into the
bulk the horizon of an object with horizon size $r$ on the brane will be 
penetrating. This can be obtained from examining the
Gregory-Laflamme instability for a black string \cite{CHR}. The result is
again given by (\ref{distance}).
Hence, physical scales $r\gg k^{-1}e^{ky_0}$ on the brane 
correspond to
distances in the transverse space which reach out into the 5D Minkowski
portion of the space in the GRS model.

Fortunately, recent developments in string theory \cite{Juan} suggest
the tools we need to do the averaging which determines the
effective theory on the brane at large distances. The technology is
known as the holographic RG 
\cite{SW,PP,Verlinde,Gubser,Jan,Hawking}. 
The idea is that integrating out short distance degrees-of-freedom
up to a four-dimensional length scale $\tilde r$,
according to a brane observer, gives rise to an effective 
theory that is described by a theory with the position of the brane 
shifted in the transverse space to $\tilde y$, where
\EQ{
\tilde r=\int^{\tilde y}_{-\infty} e^{A(y)/2}\,dy\ .
\label{cuttoff}
}
(In the above, we are assuming that $A(y)=2k|y|$, the AdS form, for $y<0$,
and so when the brane is at $y=0$ the cut-off is $k^{-1}$.) 
This is illustrated in Figure 1.
\begin{figure}
\begin{center}
\includegraphics[height=5cm]{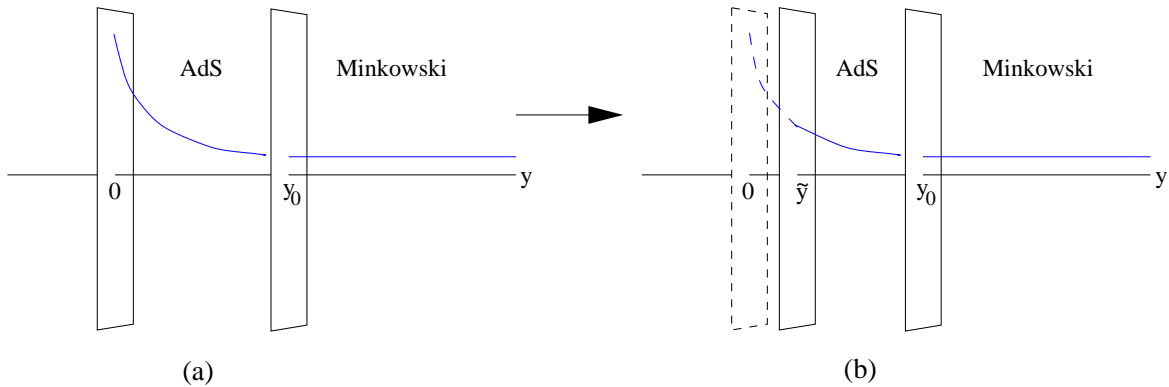}
\end{center}
\caption{\small (a) The GRS brane set-up and warp factor.  (b) Brane 
translation associated with the holographic RG running.}
\label{fig:holo1}
\end{figure}
More precisely, we cut out 
 the region $-\tilde y\leq y\leq\tilde y$ and
re-glue the two portions as illustrated in Figure 2.
The effective theory is then described by a metric
\EQ{
d\tilde s^2=dy^2+e^{-\tilde A(y)}\eta_{\mu\nu}dx^\mu\,dx^\nu\ ,
\label{effmet}
}
where 
\EQ{
\tilde A(|y|)=A(|y|+\tilde y) \ ,
}
and we have shifted $|y|\to |y|-\tilde y$ so that the effective brane
remains at $y=0$.  
Notice that in the effective background \eqref{effmet} the negative
tension branes of the GRS model, or the regions where they are
smoothed, lie at a smaller proper distance $k^{-1}e^{k(y_0-\tilde y)}$
from the effective (renormalized) brane.
Notice also that for an AdS background, the effective
metric \eqref{effmet} is identical to the original metric, 
expressing the fact that this situation describes a fixed-point of
the RG flow. 

 \begin{figure}
\begin{center}
\includegraphics[height=6cm]{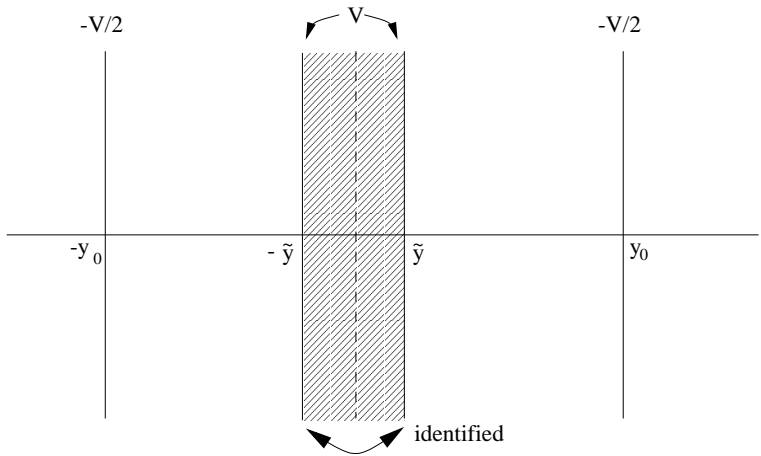}
\end{center}
\caption{\small The holographic RG: the Planck brane
is translated to $\tilde{y}$ and identified with the image brane at 
$-\tilde{y}$.}
\label{fig:holo2}
\end{figure}

In general, the effective brane theory will have a renormalized tension,
unless the background is exactly dS$_5$, AdS$_5$ or 5 dimensional Minkowski 
space.
In addition to the renormalized brane tension, there could be additional
operators induced in the effective brane action, encoding the effects
of the metric fluctuations integrated over the strip.\footnote{In addition to 
the effects we are
describing there are purely quantum renormalization effects which
are discussed in ref. \cite{Hawking}.} This is familiar from
ordinary RG running:q
when relevant operators are not present in the underlying (UV) theory, 
they can
still be generated in the effective (IR) theory. The presence of
these induced operators will ensure that the infrared physics 
is kept constant.\footnote{The first version of this paper was missing 
these induced operators, and thus neglected some important effects due
to fluctuations of the 
graviton and radion modes.}
First we discuss the renormalization of the effective brane tension. This is
simply obtained from the Israel junction condition
at the position of the effective brane:
\beq
\left[{{\partial}\over{\partial y}} g_{ij}\right] \Big|_{\tilde y} 
= - {2{\kappa^2}\over{3}} g_{ij}
 \widetilde V(\tilde y)~,
\eeq
which implies,
\EQ{
\widetilde V(\tilde y)=\frac{3A'(\tilde y)}{\kappa^2}\ .
\label{rt}
}
One can obtain the same result on the effective brane tension
by explicitly integrating the action over the strip between $0$ and 
$\tilde{y}$, and requiring that the contribution of the renormalized brane
tension to the effective 4D cosmological constant matches the combined 
contributions of the original brane tension and the curvature of the strip
we have integrated over. It is easy to see explicitly how this works in the 
general 
background metric \eqref{nmetric}.  The action is given by
\beq
\label{action}
S=- \int dy d^4x \Big[ \sqrt{g}\left(\frac{1}{2\kappa^{2}} R + \Lambda (y)
\right)\Big] -
\int dy d^4x\sqrt{\tilde{g}} \Big[ {\cal V}(y) +V \delta(y)\Big]~,
\eeq
where $\tilde{g}$ is the induced four dimensional metric at constant $y$, 
and $\Lambda (y)=-\frac{3}{2}
\kappa^{-2} A'(y)^2$, ${\cal V}(y)=\frac{3}{2}\kappa^{-2} A''(y)$ (away
from the branes) are the
static source terms \cite{US} needed to produce the metric \eqref{nmetric}.
The source term ${\cal V}$ corresponds to a ``smeared brane tension'', not 
to be confused with the actual tension $V$ or the effective tension
$\tilde{V}(\tilde{y})$.  Note 
that with these conventions Einstein's equation is given by
$G_{ab}=\kappa^2( \Lambda (y) g_{ab}+({\cal V}(y) +V\delta (y)) 
g_{\mu \nu} \delta^\mu_{a}
\delta_b^\nu )$. Our renormalization procedure corresponds to performing 
the integral in the action \eqref{action} over a slice between $0$ and 
$\tilde{y}$, and replace this space with an effective brane with tension
$\tilde{V} (\tilde{y})$. Thus the definition of the
effective brane tension is 
\begin{equation}
-\int_0^{\tilde{y}} dy \Big[ \sqrt{g} \big( \frac{R}{2\kappa^2} +\Lambda (y)
\big)
+\sqrt{\tilde{g}} {\cal V}(y) 
\Big] +\tilde{V} (0) = \sqrt{\tilde{g}(\tilde{y})}
\Big( -\frac{R^{(sing)}(\tilde{y})}{2\kappa^2}-\tilde{V} (\tilde{y})\Big),
\end{equation}
where $R^{(sing)}(\tilde{y})$ is the singular piece in the  curvature at 
the position of the effective brane. Using the expression for the
curvature $R=-\frac{10}{3} \Lambda (y)-\frac{8}{3} {\cal V}(y) -\frac{8}{3} 
\tilde{V} (\tilde{y}) \delta (y-\tilde{y})$, we find the 
formula for $\tilde{V} (\tilde{y})$ given in  \eqref{rt}.

Let us now discuss the operators induced in the effective
brane action. In a general background,
there are a variety of terms. The simplest of these is a 
four-dimensional induced Einstein-Hilbert term of the form,
\begin{equation}
\int d^4 x \sqrt{g^{ind}} R^{(4)},
\label{r4}
\end{equation}
where $g^{ind}$ is the induced metric at the 
effective brane, and $R^{(4)}$ 
is the curvature scalar calculated from the induced metric.
One way  to obtain the GRS model is as a limit of 
the two-strip model introduced in \cite{PRZ}. This 
model is just like the GRS model with two branes, except that the second brane
tension is chosen to be smaller in magnitude than $-V/2$; therefore the 
space past 
the second brane is also $AdS_5$, except with a different cosmological 
constant. This model has a localized graviton, and is qualitatively very 
similar to the RS model, except that there is an additional radion mode
appearing in the theory. We can calculate the effective brane action
in the two-strip model, and take  the limit where the 
second brane tension goes to $-V/2$. 
Following 
\cite{CGR,PRZ} we write the metric in the form, 
\begin{equation}
ds^2=e^{-A(y)}\left( 1+B(y)\,f(x)\right)
\left(\eta_{\mu\nu}+h_{\mu\nu}(x,y)+2\epsilon(y)\partial_\mu\partial_\nu
f(x)\right)\,dx^\mu\,dx^\nu+\left(1-\frac{2B'(y)}{A'(y)}f(x)\right)\,dy^2.
\label{eq:PRZmetric} \end{equation}
Note that $B'(y)/A'(y)$ is a smooth function,
even though $A'(y)$ and $B'(y)$ are
discontinuous.
Here, $h_{\mu\nu}$ are the four-dimensional 
graviton fluctuations including the zero mode
of the two-strip model
and the Kaluza-Klein modes, $f$ is
the radion, and in the GRS model the background warp factor 
$A(y)=2k|y|$ for $|y|<y_0$ and $A(y)=2ky_0$ for $|y|>y_0$. 

In order to calculate the coefficient of the 
induced operator (\ref{r4}), we will approximate the 
the graviton fluctuation in (\ref{eq:PRZmetric})
 by the zero mode $h_{\mu \nu}(x,y)=
h_{\mu \nu}(x)$. This will be a good approximation
as long as the width of the resonance is small.
Expanding the 5D bulk curvature term $\sqrt{g} R$ to zeroth order in $f$
we find it contains the
operator $e^{-A(y)} \sqrt{\tilde{g}} \tilde{R}^{(4)}$, where
$\tilde{g}$ and $\tilde{R}^{(4)}$ correspond to the 4D metric
$\tilde{g}_{\mu\nu}=\eta_{\mu\nu} + h_{\mu\nu}$. 
Thus we can integrate this term over the strip from $0$ to $\tilde y$,
to obtain the coefficient of the 4D  Ricci scalar part of the effective brane
action, $
2 \int_0^{\tilde{y}} e^{-A(y)} dy.$
If we wish to write this operator in the effective brane
action in terms of the induced metric $g^{ind}$,
then we have
\begin{equation}
S_{Ricci \,{\rm eff}}=2  
\int_0^{\tilde{y}} e^{-A(y)} dy \int d^4x \sqrt{\tilde g} 
\tilde{R}^{(4)}=
2 e^{A(\tilde{y})} \int_0^{\tilde{y}} e^{-A(y)} dy \int d^4x \sqrt{g^{ind}} 
R^{(4)}.
\end{equation}

However, this is not the full story.  In the GRS model (and the two strip
model in general) the radion is relevant in the infrared, 
and induces additional operators in the effective brane action.

\section{The Radion in the GRS Model}

In this section we study the induced 
radion kinetic term on the effective brane, following
the work of Pilo, Rattazzi and Zaffaroni \cite{PRZ}. We start with the
action
\begin{equation}
S=\int\sqrt{g}\,(M_*^3 R-\Lambda(y))\,d^5x
-\sum_i\int\sqrt{g^{ind}_i}\,V_i\,d^4x,
\end{equation}
where the bulk integral can be split into
two  regions $(0,y_0)$ and $(y_0,\infty)$ (where $y_0$ is the position of the
negative tension brane) with cosmological constants
$\Lambda_1$ and $\Lambda_2$ respectively, 
and the brane integrals involve the brane tensions $V_i$ and
induced metrics $g^{ind}_i$ on the two branes.

 The form for $B(y)$ in the metric (\ref{eq:PRZmetric}), 
in the region between the positive and
negative tension branes is 
determined from the linearized Einstein equations by insisting that the 4D
graviton fluctuations decouple from the radion.
The four-dimensional 
components of the linearized Einstein equations in the bulk are 
proportional to,
\begin{equation}
e^{2ky}(2k\,B(y)-B'(y))+2k(-4k\,\epsilon'(y)+\epsilon''(y))=0,
\label{eq:B1}\end{equation}
which is solved by \cite{PRZ}, 
\begin{equation}
B(y)=2\left[\alpha e^{2ky}+k\,e^{-2ky}\,\partial_y\epsilon(y)\right].
\label{eq:B1'}\end{equation}
The coefficient $\alpha$ is arbitrary, 
but in order
for the radion coupling to matter \cite{CGRT,CGK}
to be normalized the same as the longitudinal 
graviton we  choose it to be $\alpha=1$.
In the region beyond the negative tension brane, $y>y_0$, 
$B(y)$ takes the form,
\begin{equation}
B(y)=k'\,e^{-2k'y}\,\partial_y\epsilon(y), \label{eq:B2}\end{equation}
where for the GRS scenario we take the limit $k'\rightarrow 0.$  In this 
region (for any $k'$)
the radion dependent part of the metric is pure gauge and the equations
of motion are satisfied for any $f(x)$ \cite{CGR,PRZ}.  
The matching conditions at the positive and negative tension branes determine
\cite{PRZ}
\begin{equation}
B(0)=2, \ \ \  B(y_0)=2e^{2ky_0}\frac{k'}{(k'-k)}. \end{equation}
Note that $B(y_0)\rightarrow 0$  for the GRS model.

The pure gauge part
of the action does not contribute to the effective theory on the brane, so
with the above gauge choice for the metric the radion 
effective action is given completely by integrating
over the region between the positive and negative tension brane.

As described in the previous section
the fluctuation independent part of the action is made to vanish by the usual
fine tuning of cosmological constants and brane tension, which for GRS is,
$\Lambda_1=-12\,k^2M_*^3$ between the branes, $\Lambda_2=0$ beyond the 
negative tension
brane; $V_1=12\,k\,M_*^3$ and $V_2=-6\,k\,M_*^3$ are the brane tensions of the positive and
negative tension brane, respectively.

We focus on the radion kinetic terms in the action.  The calculation is
simplified as in \cite{PRZ} by expanding about the background metric, so that,
\begin{equation}
S_{\rm radion}=-\int_{-y_0}^{y_0} dy\int
d^4x\,\sqrt{g}\ \delta g^{\mu\nu}\,\left[M_*^3R_{\mu\nu}-\frac{1}{2}g_{\mu\nu}
\left(M_*^3R-\Lambda-\sum_iV_i\,\delta\left(\sqrt{g_{yy}}(y-y_i)\right)\right)
\right],
\end{equation}
where $\delta g^{\mu\nu}$ is the ${\cal O}\left(f(x)\right)$ term in 
$g^{\mu\nu}$, and the equation of motion is expanded to linear order in 
$f(x)$. 
Given the ansatz (\ref{eq:B1'}) for $B(y)$, the equations of motion identically
vanish to linear order except for the $(yy)$ component.  This follows from
the requirement that the equations of motion for $h_{\mu\nu}$ and $f$ are
decoupled.  The delta function contributions to the equations of motion
identically vanish given the ansatz (\ref{eq:PRZmetric}), which leaves a
single surface term for the radion kinetic part of the action:
\begin{equation}
S_{\rm radion}=\frac{M_*^3}{k}\int_0^{y_0}dy\int d^4x \sqrt{\tilde g}
\,3B'(y)\,f\Yfund f. \end{equation}
Separating the integral over the strip $(0,\tilde{y})$ from the remainder of
the bulk, the full action for the reduced bulk and renormalized brane is:
\begin{eqnarray}
S[\tilde{y}]=&&
\int_{|y|>\tilde y}\sqrt{g}\,(M_*^3 R-\Lambda(y))\,d^5x-
\sum_i\int\sqrt{g^{ind}_i}\,V_i\,d^4x
\nonumber \\
&&+\int d^4x\sqrt{\tilde g}\left[\frac{M_*^3}{k}(1-e^{-2k\tilde{y}})  R^{(4)}
+\frac{3M_*^3}{k}\left(B(\tilde{y})-B(0)\right)f\Yfund f \right],
\end{eqnarray}
where in the bulk integral the region $(0,\tilde{y})$ has been removed and the
brane tension of the positive tension Planck brane $V_1$ is replaced by its 
renormalized value $V_1(\tilde{y})$ (which is, however, constant in the GRS 
scenario for $\tilde{y}<y_0$). When $\tilde y \ge y_0$ there is a single 
tensionless brane in the effective theory, which is (aside from the 
ghostlike radion
contribution) the DGP model \cite{DGP3}.  This is a new type of
holographic duality between the GRS and DGP models, meaning that
the DGP model is in fact (modulo the radion) nothing else but the low-energy 
effective theory of the GRS model.

\section{The Graviton Propagator from Holographic RG in the GRS Model}

Before we calculate the renormalized graviton Green's function
in the GRS model, 
we pause to consider
the rationale behind the holographic renormalization group in more detail.
Performing a renormalization group transformation by moving branes through a
five dimensional space may at first seem bizarre to those not familiar
with recent developments in string theory. However we can arrive
at a simple understanding of why this procedure is reasonable by recalling
some simple classical mechanics.  In order to calculate the
potential due to some mass we can find an equipotential Gaussian surface
and require that the integral of the gradient of the potential over this
surface is equal to the mass up to appropriate factors (like Newton's
constant).  In a 5D setting we see that an equipotential surface of a
gravitational
source on the brane penetrates some distance into the
bulk around the brane. The ratio of this distance to the corresponding
distance on the brane depends in detail on the 5D metric. However as we look 
at longer distances on the brane, we are probing
further into the bulk.  Furthermore different energy/matter distributions
(including distributions that penetrate into the bulk) can have the same
equipotential surface.  Thus we are performing an
averaging over the short-distance propagation of the 5D gravitons.  
In GRS type geometries,
at large enough distances the equipotential surface will include part of the 
5D Minkowski space.  At sufficiently large
distances most of the interior of the equipotential surface will be
5D Minkowski and the gravitational potential will be that of 5D Minkowski 
space (ignoring the radion for now), falling like
the inverse of the distance squared.  Another way of saying this is that
at distances much larger than the proper distance at which the geometry becomes
flat, $\sim k^{-1}e^{ky_0}$, we
would not be able to resolve the small slice of AdS$_5$ and we would effectively
see a tensionless brane in 5D Minkowski space. However, this does not
automatically imply that the theory on the brane will be 5D Minkowski
gravity. This is because there are additional operators in the
effective action on the 
tensionless brane, which can have important consequences. For example,
the scalar anti-gravity behavior noticed in \cite{GRS2} is entirely 
due to the radion field which is localized on the tensionless brane as
envisioned in \cite{PRZ}, and does not probe the bulk.

We now discuss the long distance behavior of the GRS 
model from the point of view of the holographic picture explained above. 
We have seen that 
at the length scale $k^{-1} e^{k y_0}$  the holographic effective theory 
(corresponding to integrating out
$\tilde y$ to that scale) is 
described by a tensionless brane in a 5D Minkowski space, with the action
\begin{equation}
\int d^5 x M_*^3 e^{-2 k y_0}
\sqrt{g} R +\int d^4 x \left( M_*^3 2 b \sqrt{\tilde g} R^{(4)}
- c f \Yfund f \right),
\label{effectiveaction}
\end{equation}
where
\beq
b={{1-e^{-2 k y_b}}\over{2 k}},\ \ c=\frac{6 M_*^3}{k}. \label{eq:b,c}
\eeq
In the first term of (\ref{effectiveaction}) the factor $e^{-2ky_0}$ has been
scaled out of the metric 
so as to restore the canonical flat 5D background metric $g_{\mu\nu}=
\eta_{\mu\nu}$.
This effective theory is nothing other than 
the model recently proposed by Dvali, Gabadadze and Porrati \cite{DGP3},
in which there is a tensionless brane in 5D Minkowski space with an additional 
induced 4D curvature term, but with the ghost-like radion field necessary
to cancel the effects of the extra graviton polarization 
automatically present on the brane.
In terms of their notation the effective 5D Planck scale is
given by
\beq
\tilde{M}^3= e^{-2ky_0} M_*^3,
\eeq
and the effective 4D Planck scale is given by
\beq
M_{Pl}^2=\frac{M_*^3}{k} (1-e^{-2ky_0}).
\eeq
In \cite{DGP3} it was shown that
the graviton propagator of this theory (without the radion)
is approximately that of an almost massless 4D graviton up to distance scales
of order $\tilde{r}_0 = \frac{M_{Pl}^2}{\tilde{M}^3}$, and after that the
theory becomes 5D gravity. With our parameters, $\tilde{r}_0= k^{-1} 
e^{2k y_0}$,
but since this is already the effective theory, this scale $\tilde{r}_0$ 
is related to the original scales on the brane by 
$r_0= \tilde{r}_0 e^{k y_0}$, therefore we recover the results of 
\cite{CEH2,GRS2,PRZ}: the GRS model 
reproduces ordinary 4D Einstein gravity up to the scale 
$r_0= k^{-1} e^{3ky_0}$, and after that scale the radion will dominate and
give 4D scalar antigravity (and the graviton will contribute as in ordinary
5D Minkowski gravity).
This holographic approach thus sheds light on the
appearance of the somewhat mysterious scale
$k^{-1}e^{3 k  y_0}$.

We now show the detailed calculation of the renormalized propagator 
in the GRS model. 
Consider renormalizing from the Planck brane to 
a renormalized effective brane at the position of the negative tension
brane $y_0$.
The induced brane action for the graviton on the renormalized brane is
given by the second term in Eq. (\ref{effectiveaction})
To find the Green's function in the effective theory we follow the
method of
Giddings, Katz, and Randall \cite{GKR}.
The equation for the graviton Green's function
in transverse-traceless gauge is (after Fourier transforming from
brane coordinates to the brane momentum $p$ and taking $p^2=-q^2$, and setting 
$M_*=1$ for simplicity):
\beq
(\partial_y^2 - 2 A' \partial_y  +e^A q^2+ 2 e^{2A} b q^2 \delta(y-y_0)) 
\Delta(q,y,y') = e^{2A} \delta(y-y')
\eeq
This can be put in self-adjoint form by rescaling $\Delta$
by metric factors:
\beq
\Delta(q,y,y') = e^{A(y)+A(y')} \hat \Delta(q,y,y')~.\label{eq:rescale}
\eeq
We then have
\beq
(\partial_y^2 +A''- ( A')^2  +e^A q^2  + 2 e^{2A} b q^2 \delta(y-y_0)) 
\hat \Delta = \delta(y-y')
\label{greeneq}
\eeq
For $y,y'\ge y_0$ the Green's function can be constructed by patching
together the
solutions of the corresponding homogeneous
equation with $y<y'$ and $y>y'$, which we refer to as $\Delta_<$ and
$\Delta_>$ respectively:
\beq
\hat \Delta = \theta(y-y') \Delta_> + \theta(y'-y) \Delta_<
\eeq
Plugging the patched solution into Eq. (\ref{greeneq}) for $y'\ne y_0$
yields:
\beq
\Delta_<|_{y=y'} = \Delta_>|_{y=y'}
\label{patch1} \\
\partial_y(\Delta_>-\Delta_<)|_{y=y'}=1 
\label{patch2}
\eeq
Since we are interested in
the graviton Green's function and not just the scalar Green's function, we 
will impose the boundary condition implied by
the Israel jump condition which relates the jump in 
the derivative of the metric to the brane tension. For the 
linearized  fluctuation $h_{ij}$ around the background (in the absence of
a Ricci scalar term in the brane action)
we have 
\beq
\left[{{\partial}\over{\partial y}} h_{ij}\right]|_{y=y_0} = -2 A'(y_0)h_{ij}
\eeq
Taking account of the induced Ricci scalar term
in the effective brane action we find that the correct boundary condition is:
\beq
\partial_y \Delta_<|_{y=y_0}=-(A'+ e^{2 A} b)\Delta_<|_{y=y_0}~.
\label{bc}
\eeq
In that case the implied discontinuity in the derivative of the 
Green's function at $y=y_0$ is just what is required by Eq. 
(\ref{greeneq}) due to the 
$\delta$ function piece of $A''$ and the induced Ricci scalar.

Setting $y'=y_0$ in Eqs. (\ref{patch1}) and (\ref{patch2}) and combining
with Eq. (\ref{bc})  yields
\beq
\partial_y\Delta_>|_{y=y_0}+(A' + e^{2A} b q^2)\Delta_>|_{y=\tilde y_0} =1
\label{norm}
\eeq
as the boundary condition at the renormalized brane,
which we can use to determine the normalization of the homogeneous solution.
The solution of the homogeneous equation in the Minkowski region are just
plane waves.  The choice 
that satisfies the outgoing wave boundary condition at infinity is
\beq
\Delta_> = N e^{i q y e^{k y_0}}~,
\eeq
where $N$ is a normalization factor.

Using the normalization condition (\ref{norm}) we can determine $N$ and
find that the Green's function on the renormalized brane 
($y=y'=y_0$) is given by:
\beq
\hat \Delta(q,y_0,y_0) = {{ 1}\over
{i q e^{k y_0} +  b q^2 e^{4 k y_0} }}
\eeq
Rescaling back to the physical Green's function via (\ref{eq:rescale})
we find
\beq
\Delta(q,y_0,y_0) = {{ e^{4 k y_0} }\over
{i q e^{k y_0} +  b q^2 e^{4 k y_0} }}~,
\label{effectiveprop}
\eeq
which agrees for $q < k e^{- k y_0}$ with the full 
(and tedious) Green's function
calculation in the two-strip model with $k'\rightarrow 0$.
For $k \gg q \gg k e^{- 3 k y_0}$, using (\ref{eq:b,c}) this is approximately,
\beq
\Delta(q,y_0,y_0) \approx {{ 2 k }\over
{(1-e^{-2 k y_0}) q^2  }}~.
\eeq
which is the expected 4D propagator for intermediate scales.  Including
the 5D Planck scale couplings we can read off the effective Newton's
constant at intermediate scales to be:
\beq
G_N = {{ 2 k }\over
{(1-e^{-2 k y_0}) M_*^2  }}
\eeq
Alternatively, for small $q \ll k e^{-3 k y_0}$,
(\ref{effectiveprop}) reduces to
\beq
\Delta(q,y_0,y_0) \approx  -{{i e^{3 k y_0} }\over
{q }} +{{1}\over{2 k}} (e^{6 k y_0}- e^{4 k y_0})~.
\eeq
Thus we see that (neglecting the effects of the radion) 
at long distances the graviton Green's function goes over
to 5D behavior, with an associated gravitational potential that 
falls off like $1/r^2$.

\section{Cosmology of Theories with Quasi-localized Gravity}

We have seen in the previous section that, as predicted in \cite{GRS2,PRZ},
the GRS model (and theories with quasi-localized gravity in general) 
interpolates between ordinary 4D Einstein gravity and (due to the
presence of the radion field with negative kinetic term)  4D 
scalar anti-gravity. Since the cosmology of brane 
models 
has recently attracted a lot of attention
\cite{BDL,Kaloper,CGKT,CGS,CF,CGRT,Olive,BDEL,othercosm},
we will briefly sketch how the cosmology of the GRS model would work, even
though due the radion instability this model is probably not a realistic 
model of the Universe. First we consider the case of pure radiation
on the positive tension brane in the GRS model. Since the energy-momentum
tensor for radiation is traceless, the radion does not couple to this type
of source. Thus for pure radiation, the effect of the radion should be
negligible at the classical level. Therefore, for a radiation dominated universe,
the GRS model interpolates between the RS model at high (but below  $M_{Pl}$) 
energies and 
a tensionless brane in 5D Minkowski space at very low energies (that is at
very large distances). We expect the expansion of the Universe to be
dominated by the largest scales, since the gravitational energy itself
is dominated by the largest distances. This is so because the gravitational 
energy in a sphere grows as the square of the radius in the case of 4D gravity
(and as the radius for 5D gravity). The reason is that even though the 
potential is decreasing with the distance, there is much more matter close to
the edge of the sphere than in the middle. Therefore we will assume that 
the expansion of the Universe should be described by the effective 
holographic brane model with the cut-off given by the size of the Universe
(that is, the region in causal contact since the Big Bang).
This immediately indicates the type of expansion one expects here: as long as
the size of the Universe is smaller than the distance where Einstein gravity
turns into 4D anti-gravity, we expect the expansion to be given by 
that of a brane in AdS space, that is the ordinary Friedmann equation.
Once the size of the Universe reaches the critical distance, it will be
given by the cosmology of a tensionless brane in 5D Minkowski space.
Bin\'etruy, Deffayet and Langlois (BDL) showed that the
cosmology of a tensionless brane in Minkowski space does not
reproduce the ordinary Friedmann equations; instead it predicts a Hubble
law of the form $H^2 \propto \rho^2$, where $\rho$ is the energy 
energy density of matter on the brane \cite{BDL}. Thus, once the 
critical distance is reached, 
the expansion will change to that of BDL. Since 
in this case $H^2 \propto \rho^2$ instead of the ordinary $H^2 \propto \rho$,
the expansion will change  once the 5D phase is reached. The transition will 
not be as abrupt as in the largest-scale-dominance approximation, however, 
away from the transition region largest-scale-dominance should be a good 
approximation. The expansion equation for this case has been 
investigated in detail in \cite{CGKT}. The Friedmann equation is given by
\begin{equation}
\frac{\ddot{a_0}}{a_0}+ \left( \frac{\dot{a_0}}{a_0}\right)^2=
-\frac{\kappa^4}{36} \rho(\rho+3 p),
\label{BDL1}
\end{equation}
where $a_0(t)$ is the scale factor at the positive tension brane,
and $\rho, p$ are the radiation energy and pressure densities 
introduced on the brane as
\begin{equation}
T^{\rm brane\ \mu}_{\phantom{\rm brane\ }\nu}=b^{-1}\delta (y) \ {\rm diag}
(-\rho ,p,p,p)\ ,
\end{equation}
and the five dimensional metric (which includes both the background and the 
expansion due to the matter sources on the brane) is given by
\begin{equation}
ds^2= b^2 (y,t) dy^2+a^2 (y,t) d\vec{x}^2-n^2 (y,t) dt^2\ .
\end{equation}
In \cite{CGKT} it was shown, that there are two types of solutions 
for the above equation (\ref{BDL1}). One is the solution obtained
in \cite{BDL}, for which $a_0(t) \sim t^{\frac{1}{4}}$, while for the
other solution
\begin{equation}
a_0(t) \sim t^{\frac{1}{2}} \left( 1-\frac{\kappa^4}{36} \frac{\rho^2(t_i)
a_0(t_i)^8}{t^2} +\ldots \right).
\end{equation}
Thus for the case of pure radiation, there are two known solutions to the
expansion equations. In one solution 
the expansion of the Universe would slow down to $t^{\frac{1}{4}}$, while 
for the other solution (if the densities are small compared to the 
expansion rate at the critical distance) the solution remains an
essentially unchanged $t^{\frac{1}{2}}$ expansion law, up to small 
corrections. Which solution is actually realized depends on initial
conditions but we expect that for generic initial conditions (where
the radiation dominated universe expands normally in the intermediate
4D regime) that the second solution should hold to good approximation.

The situation will be very different for other types of matter,
for example consider an ordinary matter dominated Universe. In this case,
the expansion will again start out as an ordinary 4D expansion,
given by the power law $a_0(t) \sim t^{\frac{1}{3}}$. However, for
distances larger than the critical distance, the radion will start dominating 
the expansion. We will approximate the expansion equations for this case
by assuming that we have a tensionless brane in 5D Minkowski space, plus
the radion localized on in. At distances where the 5D nature of gravity
becomes apparent, the induced 4D curvature term can be neglected compared
to the 5D curvature term. In this case, one can derive the general 
expansion equations, which will now also involve the radion field $\varphi$.
To find the expansion equations, we start with an effective action
at very large distances of the form
\begin{equation}
\int d^5x \frac{1}{2 \kappa^2} \sqrt{g}R +\int d^4x \frac{1}{2} \partial_\mu 
\varphi \partial^\mu \varphi,
\end{equation}
where $\varphi=\sqrt{2c} f$ represents the radion field with the wrong-signed
kinetic term, and $\kappa$ here stands for the Newton's
constant in the effective holographic theory, which is 
related by the warp factor to the fundamental parameter of 
the the model. Similarly to \cite{BDL}, one can derive the expansion equation 
for this action. The difference compared to \cite{BDL} will be that
the scalar is localized on the effective brane as well, therefore they will
appear as additional delta-function like sources, which will modify
the ``jump-equations'' to
\begin{equation}
\frac{[a']}{a_0 b_0} =-\frac{\kappa^2}{3} \rho + \frac{\kappa^2}{6 n^2} 
\dot{\varphi}^2, \ \ \frac{[n']}{n_0 b_0}=\frac{\kappa^2}{3} (3p+2\rho )
+\frac{ \kappa^2}{6 n^2} 
\dot{\varphi}^2,
\end{equation}
where $[a']$ denotes the jump of the $y$ derivative of the scale factor at the position of the 
brane. Then similarly to \cite{BDL} one obtains the expansion equation by
substituting these jumps into the $55$ component of Einstein's equation.
The result is given by
\begin{equation}
\frac{\ddot{a_0}}{a_0}+ \left( \frac{\dot{a_0}}{a_0}\right)^2 =
-\frac{\kappa^4}{36} \rho(\rho+3 p) +\frac{\kappa^4}{72}\dot{\varphi}^4
+\frac{\kappa^4}{72}\dot{\varphi}^2 (\rho + 3p),
\label{BDL2}
\end{equation}
where we have rescaled time such that $n_0=n(0)=1$.
However, this is not the whole story, since $\rho$ and $p$ are also
the sources for $\varphi$, which is given by the scalar equation of motion 
on the brane\cite{CEH2,GRS2}. 
The linear coupling to matter on the brane (neglecting the 
dimension 6  derivative couplings) is given by
\beq
\varphi \frac{ \delta S}{ \delta \varphi} \large| _{\phi=0} 
= \varphi \frac{ \delta S}{ \delta (g^{ind})^{\mu \nu}} 
\frac{ \delta (g^{ind})^{\mu \nu}}{ \delta \varphi}=  
-\sqrt{g^{ind}} 
\tilde{T}_{\mu \nu}  e^{-A(0)}\tilde{g}^{\mu \nu} \frac{B(0)}{2} 
\,\frac{ \varphi(x)}{\sqrt{2c}}~.
\eeq
Note, that the matter is assumed to be at the original Planck brane,
therefore there will be no additional warp factor appearing
in the coupling.
Thus the radion equation of motion is 
\begin{equation}
\Yfund \varphi =-\frac{1}{\sqrt{2c}}(3p-\rho ),
\end{equation}
which, assuming that there is only time dependence, would give the equation
\begin{equation}
\ddot{\varphi} = \frac{1}{\sqrt{2c}} (3p -\rho ).
\label{exp}
\end{equation}
Thus for general matter the expansion is 
determined by the coupled equations
(\ref{BDL2}) and (\ref{exp}). Due to the antigravitational nature
of the interaction mediated by the radion we expect that (like
in \cite{CGKT}) the Universe would reach
a maximal size and then recollapses. 
Since the ghostlike radion is likely to make the
model unstable anyway, we will not pursue the solutions of the 
expansion equation sketched above.

\section{Conclusions}
 
We have examined a class of 5D metrics with embedded 3-branes. Along the
extra dimension these
models are asymptotically Minkowski, and gravitons are
quasi-localized on a  brane.  One would expect that at sufficiently
large
distances the details near the brane are irrelevant and effectively
there
is a tensionless brane in 5D Minkowski space (a brane with non-zero
tension would curve the space around it).  We find that a holographic
renormalization
group analysis confirms this intuitive picture for the graviton.  
The renormalization
group analysis also shows how  the radion effectively gives rise to
scalar ``anti-gravity'' at
long distances by an induced radion coupling on the brane with negative kinetic
term. In fact, at intermediate and large distances the holographic
effective theory is equivalent to the recently proposed model of Dvali et al.,
where the tensionless brane in 5D Minkowski space also has an induced
4D curvature term on the brane. 
Thus the behavior of quasi-localized gravity
in GRS-type models at different length scales is as follows: 
at very short distances the theory 
is five dimensional (both scalar potential and tensor structure); at 
intermediate scales it is given by ordinary 4D gravity with corrections
that can be arbitrarily small; and at ultra-large distances the graviton
is again five dimensional and the 4D radion dominates. 
Thus these models do not seem to be
internally consistent; however, if a generalized model could
eliminate the radion from the light degrees of freedom (as can happen
in RS models) they might  produce viable
cosmologies  which decelerate after a late epoch.  To be consistent
with current observations this epoch must be later than the current
epoch.

\section*{Acknowledgements}

We would like to thank Michael Graesser, Juan Maldacena, and Martin Schmaltz
for useful conversations, and to Gia Dvali, Gregory Gabadadze, Massimo
Porrati and Valery 
Rubakov for correspondence and for comments
on the manuscript. We also thank Ami Katz, Luigi Pilo, Lisa Randall,
Riccardo Rattazzi and Alberto Zaffaroni for comments and criticism 
on the first version of this paper, from which we have greatly benefitted. 
C.C. is an Oppenheimer Fellow at the  Los Alamos National Laboratory.
C.C., J.E. and T.J.H. are supported
by the US Department of Energy under contract W-7405-ENG-36.
J.T. is supported
in part by the NSF under grant PHY-98-02709.
J.T. thanks the T8 group at LANL for its hospitality while part of this 
work was completed.

\end{document}